# Two-dimensional superconductivity at the surfaces of $KTaO_3$ gated with ionic liquid

Tianshuang Ren[1#], Miaocong Li[1#], Xikang Sun[1#], Lele Ju[1], Yuan Liu[1], Siyuan Hong[1], Yanqiu Sun[1], Qian Tao[1], Yi Zhou[2,3,4]*, Zhu-An Xu[1,5]*, and Yanwu Xie[1,5]*

[1]Interdisciplinary Center for Quantum Information, State Key Laboratory of Modern Optical Instrumentation, and Zhejiang Province Key Laboratory of Quantum Technology and Device, Department of Physics, Zhejiang University, Hangzhou 310027, China
[2]Beijing National Laboratory for Condensed Matter Physics & Institute of Physics, Chinese Academy of Sciences, Beijing 100190, China
[3]Songshan Lake Materials Laboratory, Dongguan, Guangdong 523808, China
[4]Kavli Institute for Theoretical Sciences and CAS Center for Excellence in Topological Quantum Computation, University of Chinese Academy of Sciences, Beijing 100190, China
[5]Collaborative Innovation Center of Advanced Microstructures, Nanjing University, Nanjing 210093, China

#These authors contributed equally to this work.
*To whom correspondence should be addressed. E-mail: yizhou@iphy.ac.cn (Y.Z.); zhuan@zju.edu.cn (Z. A. X.); ywxie@zju.edu.cn (Y. X.)

**Abstract.** The recent observation of superconductivity at the interfaces between $KTaO_3$ and $EuO$ (or $LaAlO_3$) offers a new example of emergent phenomena at oxide interfaces. This superconductivity exhibits an unusual strong dependence on the crystalline orientation of $KTaO_3$ and its superconducting transition temperature $T_c$ is nearly one order of magnitude higher than that of the seminal $LaAlO_3/SrTiO_3$ interface. To understand its mechanism, it is crucial to address if the formation of oxide interfaces is indispensable for the presence of superconductivity. Here, by exploiting ionic liquid (IL) gating, we obtain superconductivity at $KTaO_3$ (111) and (110) surfaces with $T_c$ up to 2.0 K and 1.0 K, respectively. This oxide-interface-free superconductivity gives a clear experimental evidence that the essential physics of $KTaO_3$ interface superconductivity lies in the $KTaO_3$ surfaces doped with electrons. Moreover, the ability to control superconductivity at surfaces with IL provides a simple way to study the intrinsic superconductivity in $KTaO_3$.

**Introduction.**

Since the discovery of superconductivity at the LaAlO$_3$/SrTiO$_3$ (LAO/STO) interface(*1*), oxide interface superconductivity has been attracting increasing interest(*2–9*). It is two dimensional (2D) in nature(*1, 6, 7, 9–14*) and can be tuned by applying an electrical gate bias(*2, 6, 7, 14*), which provides a promising platform to explore the rich physics of 2D superconductors and relevant quantum phase transitions. In LAO/STO, the superconductivity, or more precisely, the conduction, locates in a thin STO layer near the interface(*1*), and the superconducting transition temperature $T_c$ is comparable with the electron-doped bulk STO(*15*), the delta-doped STO thin layer(*16*), and the ionic liquid (IL)-gated STO surface(*17*). And the interfacial superconductivity shows no strong dependence on the STO crystalline orientation(*1, 10, 11, 18*). However, the newly discovered superconductivity in KTaO$_3$ (KTO) interfaces(*12–14, 19, 20*), with a one order of magnitude higher $T_c$ than LAO/STO, behaves in a very different manner, although KTO shares many common features with STO(*12, 21, 22*).

While STO is the first known superconducting semiconductor since 1964(*15*), the first report of superconductivity in KTO was in 2011(*22*), when a $T_c \sim$ 50 mK was observed at KTO(001) surface gated with IL. After that, there was a decade of silence until the recent discovery on a family of KTO interfaces(*12–14, 19, 20*). Unusually, the superconductivity of these interfaces exhibits a strong dependence on the KTO crystalline orientations: the (111)(*12*) and (110)(*13*) interfaces have an optimal $T_c$ of 2.2 and 0.9 K, respectively, a big leap from aforementioned 50 mK(*22*); in contrast, no superconductivity was detected at the (001) interface down to 25 mK(*12*). It is yet to be unveiled why the EuO (or LAO)/KTO interfaces have such a dramatically enhanced superconductivity and why the superconductivity exhibits a strong dependence on the KTO crystalline orientation.

A crucial question follows: Does the EuO (or LAO) overlayer play a fundamental role in the occurrence of superconductivity or mainly serve as a way to induce charge carriers(*20*), say, electrons? To address this issue, it is in high demand to explore the electron-doped KTO surface that is not capped with any oxide layer. For this purpose, IL gating is an ideal tool, in which high-concentration electrons can be induced by applying a gating bias, whose underlying mechanism could be electrostatic charge modulation(*17*), oxygen vacancy formation(*23*) or hydrogen insertion(*24*). In this work, we report our IL-gating experiments on KTO surfaces of all the three principle crystalline orientations. A strong crystalline-orientation-dependent 2D superconductivity, which is essentially the same as that observed at the EuO (or LAO)/KTO interfaces, is achieved on KTO surfaces.

**Results**

**Device fabrication**

The device fabrication is illustrated sequentially in Fig.1 (A-E). Using standard optical lithography and lift off techniques, we confined a clean and flat KTO surface (Fig. 1A and fig. S1) to a Hall-bar region by coating the other area with a 200-nm-thick amorphous AlO$_x$ (a-AlO$_x$) layer (Fig. 1B). The gating electrodes of a combined metal layer of Ti(10 nm) and Au(50 nm) were deposited on the a-AlO$_x$ layer (Fig. 1C). Then we deposited a thin amorphous LAO (a-LAO) layer on six contacting pads of the Hall bar (Fig. 1D). The deposition of the a-LAO layer made the underneath KTO metallically conducting (fig. S3), which ensures better electrical contacts to the central channel when gated by IL. A tiny droplet of the IL N,N-Diethyl-N-methyl-N-(2-methoxyethyl)ammonium bis(trifluoromethanesulfonyl)imide (DEME-TFSI) covered the channel and lateral gate electrodes (Fig. 1E). The gate voltage $V_G$ was applied as schemed in Fig. 1E. A photo image of a typical device is shown in Fig. 1F. More details in fabricating devices can be found in the section of **Materials and Methods**.

**Emergence of superconductivity**

We measured the transport properties of IL-gated KTO surfaces in fabricated devices. Fig. 2A shows the temperature dependence of the channel sheet resistance $R_{sheet}$ of a set of typical samples with three different crystalline orientations (111), (110) and (001), gated at $V_G = 4.5$ V. An overall metallic conduction was observed on these three types of KTO surfaces. At a given temperature, $R_{sheet}$ decreases in an order following (111), (110) and (001). As demonstrated in Fig. 2B, a superconducting state occurs at low temperatures in these (111) and (110) samples, with a mid-point $T_c$ of 1.66 K and 0.94 K respectively. In contrast, no superconducting transition was detected in the (001) sample down to 0.4 K (Fig. 2B). Hall effect measurements show that charge carriers are electrons. The sheet carrier density $n_{sheet}$ measured at 2 K for the (111), (110), and (001) samples are $4.37 \times 10^{13}$ cm$^{-2}$, $4.44 \times 10^{13}$ cm$^{-2}$, and $5.21 \times 10^{13}$ cm$^{-2}$, respectively (Fig. 2C). The presence of superconductivity is also supported by the following two facts: the zero-resistance state in the IL-gated (111) or (110) surfaces can be completely suppressed by applying a magnetic field perpendicular to the surface (Fig. 2D); a fairly well-defined critical current was observed in the voltage versus current (*V-I*) characteristics (Fig. 3A for (111) and fig. S4a for (110)).

The emergence of superconductivity is quite robust. As shown in Fig. 2G, we have observed superconductivity in multiple IL-gated KTO(111) samples. The observed maximum mid-point $T_c$ reaches 2.04 K (see (111)_B, #2), which is comparable with the optimal $T_c$ achieved at the EuO (or LAO)/KTO(111) interfaces(*12, 14*). We would like to emphasize that the interface between a-LAO and KTO (Fig. 1D and fig. S3) plays a negligible role in the emergence of superconductivity on IL-gated KTO surfaces since the superconductivity survives in samples without depositing a-LAO (see (111)_C).

**2D superconductivity**

The superconductivity at the IL-gated KTO(111) and KTO(110) surfaces exhibits 2D features. For a 2D superconductor, the superconducting transition is of the Berezinskii-Kosterlitz-Thouless (BKT) universality class. At the vortex-antivortex pair unbinding transition, the current causes the proliferation of vortices, resulting in a $V \propto I^\alpha$ power-law dependence, with $\alpha(T_{BKT}) = 3$. As shown in Fig. 3B and fig. S4b, for samples gated at $V_G$ = 4.5 V, the power-law dependence was clearly demonstrated; the BKT transition temperature $T_{BKT}$ is identified to be 1.57 K for (111) (Fig. 3B) and 0.81 K for (110) (fig. S4b) respectively. We also derived the $T_{BKT}$ from the $R_{sheet}(T)$ characteristics (see **Materials and Methods**), which gives rise to $T_{BKT}$ = 1.69 K for (111) (Fig. 2E) and $T_{BKT}$ = 1.00 K for (110) (Fig. 2F). These $T_{BKT}$ values, and the mid-point $T_c$ derived from the $R_{sheet}(T)$ curves, agree fairly well with each other.

The 2D nature of the IL-gated KTO surface superconductivity is further revealed by the large anisotropy in magnetoresistance in magnetic fields applied perpendicular and parallel to the KTO surface. A typical set of temperature-dependent magnetoresistance data for the KTO(111) surface gated at $V_G$ = 4.5 V is shown in fig. S5, and the upper critical field $\mu_0 H_{c2}$ derived from it is shown in Fig. 3C. From the temperature dependences of $\mu_0 H_{c2}^{\perp}(T)$ and $\mu_0 H_{c2}^{\parallel}(T)$, we extract the zero temperature Ginzburg-Landau coherence length $\xi_{GL}(0)$ and the corresponding superconducting layer thickness $d_{sc}$ to be ~24.7 nm and ~9.4 nm, respectively (see **Materials and Methods**). The same measurements and analysis on the KTO(110) surface gated at $V_G$ = 4.5 V (see fig. S6) give rise to $\xi_{GL}(0) \sim 45.8$ nm and $d_{sc}$ ~17.4 nm. It is noted that, in both (111) and (110) samples, $d_{sc} < \xi_{GL}(0)$ is consistent with 2D superconductivity. In addition, the $\xi_{GL}(0)$ values are comparable with, while the $d_{sc}$ values are larger than, those of the corresponding EuO (or LAO)/KTO interfaces(*12–14*).

**$V_G$ dependence of transport properties**

Finally, we examine the $V_G$-dependence of transport properties on both IL-gated (111) and (110) surfaces. For the (111) surface, by sweeping $V_G$ from 4.5 V to 1.75 V, we observed superconductivity at $V_G \geq 2.5$ V (Fig. 4A), and the mid-point $T_c$ decreases as $V_G$ is lowering (Fig. 4B). The sample turned into a metallic or weakly insulating state at $V_G \leq 2.0$ V, and became too insulating to be measured at $V_G < 1.75$ V. Similar observations were made on the (110) surface except for a larger onset $V_G$ for superconductivity (3.5 V on (110) versus 2.5 V on (111)) (Fig. 4E). Interestingly, for both surfaces, in the superconducting range the $V_G$ modulation of $n_{sheet}$ is only a minor effect and does not follow that expected from an electrostatic tuning. For example, for the (111) surface, the $n_{sheet}$ measured at 2 K increases from $4.44 \times 10^{13}$ cm$^{-2}$ to $5.48 \times 10^{13}$ cm$^{-2}$ when $V_G$ changes from 4.5 V to 2.5 V (Fig. 4C). This suggests gating effect is not purely electrostatic. It is noted that we never detected significant difference in surface morphology (fig. S2) and crystalline structure (fig. S7) in samples before and after IL gating experiments. Meanwhile, the gating-induced conduction vanished when IL was removed. Thus, as well as the electrostatic effect, we cannot attribute the gating effect to the IL-gating-induced oxygen vacancies or other electrochemical processes solely. It is mostly likely a hybridized process.

Further insights can be gained from the $V_G$-dependence of the Hall mobility $\mu_{Hall}$ (Fig. 4D). As $V_G$ is lowering and approaching the quantum phase transition point on which the superconductivity is completely suppressed, the Hall mobility $\mu_{Hall}$ increases notably. Thus, it is reasonable to speculate that the mobility $\mu_{Hall}$, rather than the charge carrier density $n_{sheet}$, plays a major role in the tuning of KTO surface superconductivity.

**Summary**

In summary, we have demonstrated that a strong crystalline-orientation-dependent 2D superconductivity can be induced on KTO surfaces by IL gating technique. The very similarities between this surface superconductivity and its cousin, i.e., the one observed in the interfaces between KTO and other oxides (EuO or LaAlO$_3$), provide a clear evidence that the essential physics of KTO interfacial superconductivity lies in the KTO surfaces doped with electrons. As a last remark, we comment that, the present and previous experiments(*12*, *14*) indicate that in a given KTO sample a relatively low mobility favors the superconductivity, which could be a clue to further tackle the unusual KTO superconductivity.

**Materials and Methods**

**Device fabrication.** The commercial 0.5 mm-thick KTO single-crystalline substrates were purchased from Hefei Kejing Materials Technology Co., Ltd. Hall-bar devices were pre-patterned onto the surface of the KTO substrates using standard optical lithography and lift off techniques. The ~200-nm-thick a-AlO$_x$ hard mask (blue regions, Fig. 1B) was grown by pulsed laser deposition at room temperature under 0.01 mbar oxygen atmosphere. The interface between a-AlO$_x$ and KTO is highly insulating, and thus is inactive. The active uncovered KTO surface is confined within the Hall-bar area (red regions, Fig. 1B). The width of the central channel is 100 μm for Samples (111)_A, (110)_A, and (001)_A, and 20 μm for Samples (111)_B and (111)_C. The gate electrodes (yellow regions, Fig. 1C) were made by a combined bilayer of 10-nm titanium and 50-nm gold metals, both of which were evaporated by electron-beam evaporation. A 20-nm (or 10-nm for (111)_B) a-LAO layer was deposed on the Hall-bar contacting pads (grey regions, Fig. 1D). The deposition was made by pulsed laser deposition at room temperature, in $2\times10^{-7}$ mbar background vacuum. Such an interface between a-LAO and KTO is metallic and thus facilitates the electric contacts to the central channel when gated with IL. Before putting IL, the sample surface was treated by oxygen plasma for 20 seconds to remove any remaining photoresist.

**Atomic force microscopy (AFM).** The AFM data were taken using non-contact mode on a Park NX10 system.

**X-ray diffraction (XRD).** The XRD data were taken using a monochromated Cu-Kα source on a 3-kW high-resolution Rigaku Smartlab system.

**Electrical contacts and transport measurements.** The electrical contacts to the devices were made by ultrasonic bonding with Al wires. The transport measurements were carried out in a commercial DynaCool Physical Property Measurement System (PPMS, Quantum Design) with a 3He insert. A DC current measurement method was used.

**Ionic liquid gating.** The electrolyte we used was N,N-Diethyl-N-methyl-N-(2-methoxyethyl)ammonium bis(trifluoromethanesulfonyl)imide (DEME-TFSI, Kanto Chemical Co., Inc). For accumulating carriers, the gate voltage $V_G$ was applied at 250 K using a Keithley 2611B source meter. For changing $V_G$, we set $V_G$ to 0 in low temperature, warmed up the devices to 300 K, and then cooled them down to 250 K where the new $V_G$ was applied.

**Derivation of $T_{BKT}$ with $R_{sheet}(T)$ characteristics.** The $R_{sheet}(T)$ data were fitted using $R_{sheet}(T) = R_0 \exp[-b(T/T_{BKT}-1)^{-1/2}]$, where $R_0$ and $b$ are material parameters(25).

**Analysis using Ginzburg-Landau form.** The Ginzburg-Landau coherence length, $\xi_{GL}$, is extracted using the linearized Ginzburg-Landau form(16, 26) $\mu_0 H_{c2}^{\perp}(T) = \frac{\phi_0}{2\pi \xi_{GL}^2(0)}(1-\frac{T}{T_c})$, where $\phi_0$ is the flux quantum and $\xi_{GL}(0)$ is the extrapolation of $\xi_{GL}$ to $T = 0$ K. For a 2D superconductor, $\mu_0 H_{c2}^{\parallel}(T) = \frac{\phi_0 \sqrt{12}}{2\pi \xi_{GL}(0) d_{SC}}(1-\frac{T}{T_c})^{1/2}$, where $d_{sc}$ is the superconducting layer thickness(16, 26).

**Acknowledgments**

This work was supported by the National Natural Science Foundation of China (11934016, 12074334, 11774305, 12034004, 11774306), the National Key R&D Program of China (2017YFA0303002, 2019YFA0308602), the Key R&D Program of Zhejiang Province, China (2020C01019, 2021C01002), the Strategic Priority Research Program of Chinese Academy of Sciences (No. XDB28000000), and the Fundamental Research Funds for the Central Universities of China.

**Figures**

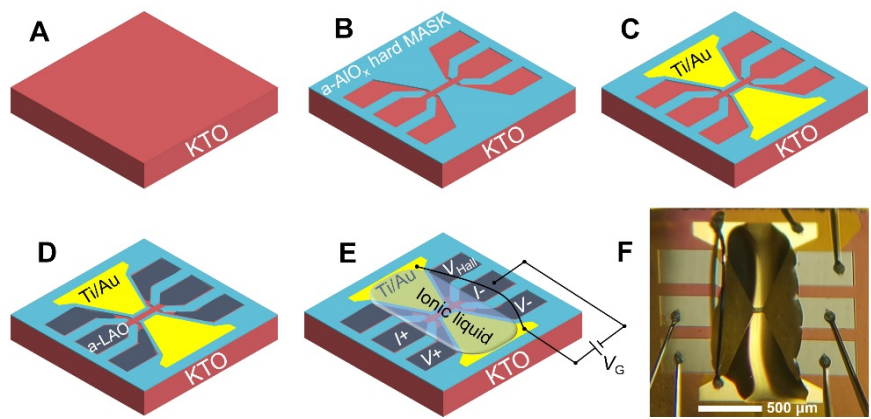

**Fig. 1. Schematic of gating KTO surfaces with ionic liquid (IL). (A-E)**, Step-by-step processes for fabricating devices. (**A**) Clean KTO single-crystalline substrate with flat surface. (**B**) Confining KTO surface to a Hall-bar region by coating the other area with 200-nm-thick a-AlO$_x$ layer as a hard mask. The a-AlO$_x$/KTO interface is highly insulating. (**C**) Depositing Ti(10 nm) and Au(50 nm) metallic layers to form gating electrodes. (**D**) Deposing a thin a-LAO layer on six contacting pads of the Hall bar. The deposition was performed in vacuum, at room temperature. This process metallizes the underneath KTO surface and thus improves the electric contact to the central Hall-bar channel when gated with IL. (**E**) Putting a tiny drop of IL (DEME-TFSI) on the central area as schemed. Gating configuration and electrical contacts to the device are as labeled. Before putting the IL, the device surface was cleaned by oxygen plasma to remove any residual photoresist. (**F**) Photo image of a typical device with bonded Al contacting wires.

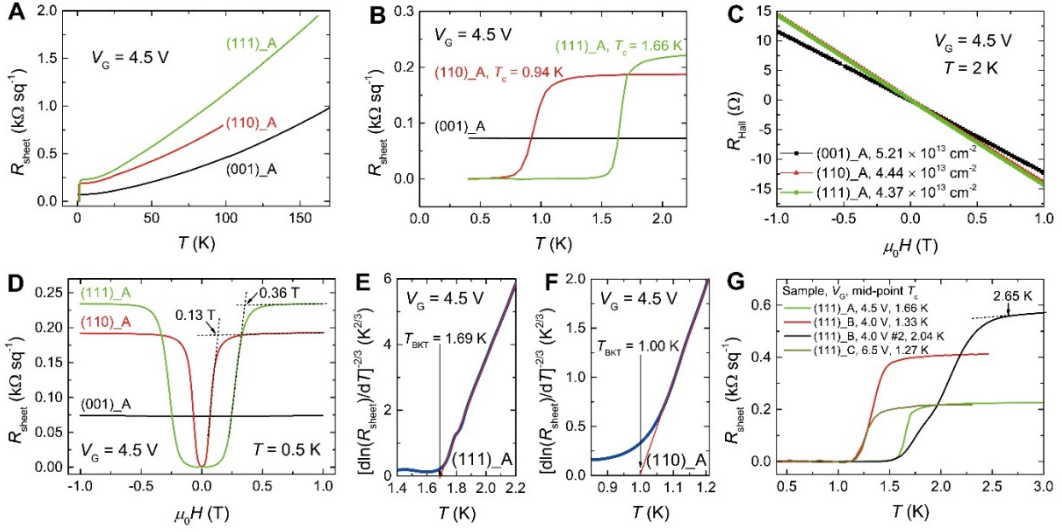

**Fig. 2. Transport properties of KTO surfaces gated with IL.** (**A**) Sheet resistance $R_{sheet}$ as a function of temperature for a set (denoted by A) of typical samples with three different crystalline orientations, at $V_G$ = 4.5 V. (**B**) A close view of $R_{sheet}(T)$'s at low temperatures. (**C**) Hall resistance $R_{Hall}$ as a function of magnetic field at $T$ = 2 K. (**D**) $R_{sheet}$ as a function of a magnetic field applied perpendicular to the surface, measured at $T$ = 0.5 K. (**E, F**) The same $R_{sheet}(T)$ dependences of (111)_A and (110)_A shown in (**B**), plotted on a $[d\ln(R_{sheet})/dT]^{-2/3}$ scale. Solid lines indicate the scaling law for a BKT transition. (**G**) $R_{sheet}(T)$ for multiple gated KTO(111) samples (denoted by A, B and C) showing resistive superconducting transitions. The $V_G$ and mid-point $T_c$ for each sample are as labeled. The optimal $T_c$ was observed in sample (111)_B when gated and measured in a second run (#2). Note that the process 1(**D**) (depositing a-LAO) was skipped in fabricating (111)_C.

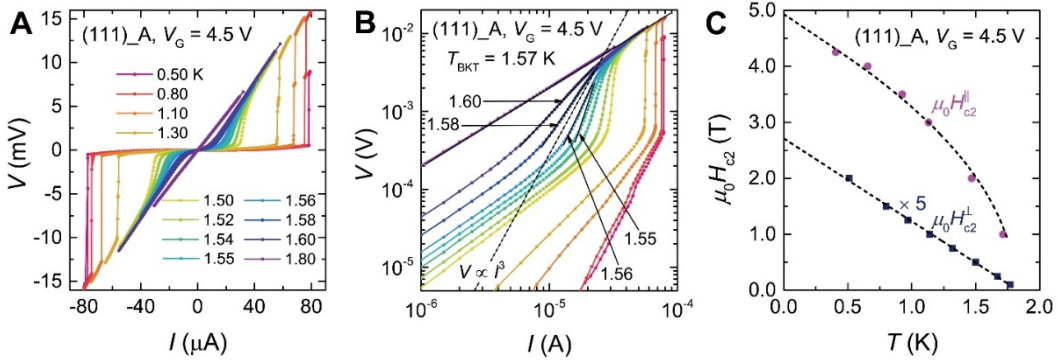

**Fig. 3. Features of 2D superconductor for the IL-gated KTO(111) surface.** (**A**) Temperature-dependent voltage-current (*V-I*) characteristics of the (111)_A sample at $V_G$ = 4.5 V. The channel width is 100 μm. (**B**) the same *V-I* curves on a logarithmic scale. The long dashed line corresponds to $V \propto I^3$ dependence and shows that 1.56 K < $T_{BKT}$ < 1.58 K. (**C**) The upper critical field $\mu_0 H_{c2}$ versus temperature. The $\mu_0 H_{c2}^\perp$ (squares; field data multiplied by 5 for clarity) and $\mu_0 H_{c2}^\parallel$ (circles) label the upper critical field perpendicular and parallel to the sample surface respectively. Dashed lines are fits to linearized Ginzburg-Landau theory.

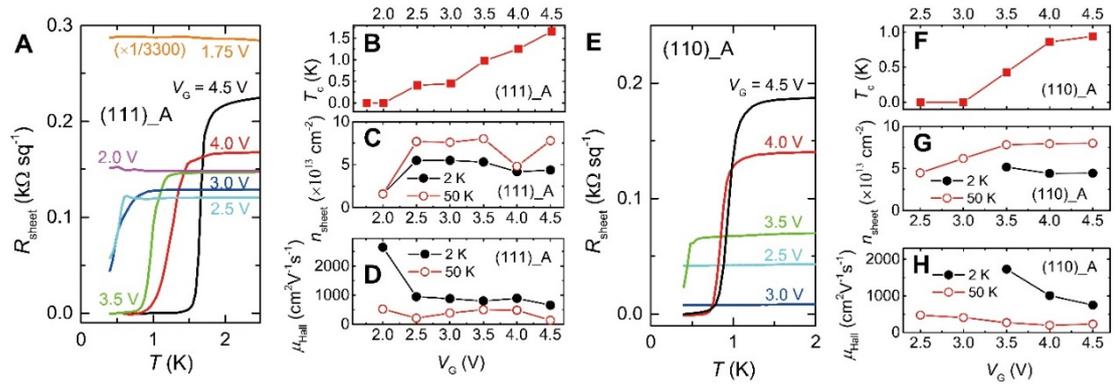

**Fig. 4. Gate voltage $V_G$ dependence of transport properties.** (**A-D**) (111) and (**E-H**) (110) surfaces. (**A**, **E**) Temperature dependence of $R_{sheet}$ at different $V_G$s. (**B**, **F**) Mid-point $T_c$, (**C**, **G**) sheet carrier density $n_{sheet}$, and (**D**, **H**) Hall mobility $\mu_{Hall}$ at

# Supplementary Materials for

# Two-dimensional superconductivity at the surfaces of KTaO$_3$ gated with ionic liquid

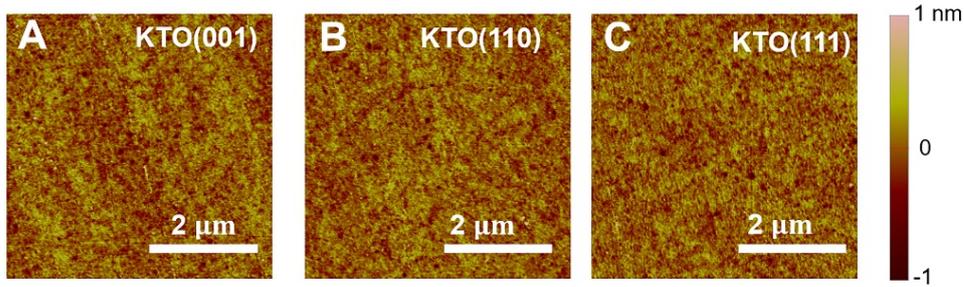

**Fig. S1. Atomic force microscopy images of three typical KTO substrate surfaces.**
(**A**) KTO(001), (**B**) KTO(110), and (**C**) KTO(111). The root-mean-square roughness over the whole 5 μm by 5 μm area is ~0.21 nm, ~0.18 nm, and ~0.18 nm for (**A**), (**B**), and (**C**), respectively.

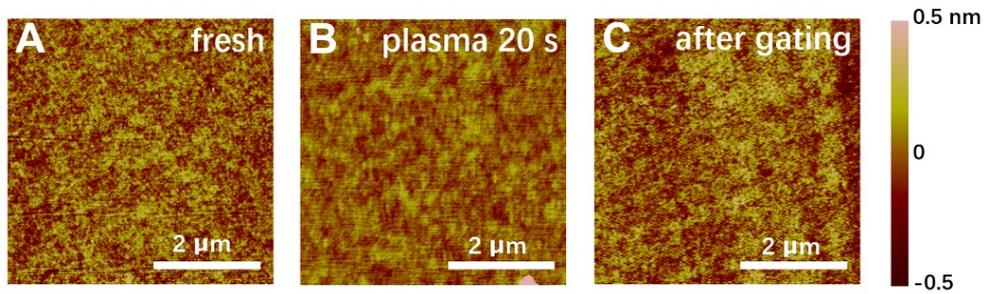

**Fig. S2. Atomic force microscopy images of a KTO(111) substrate surface in three different situations.** (**A**) Fresh surface. (**B**) After treating the surface with oxygen plasma for 20 seconds. (**C**) After an IL-gating experiment and removing IL. The gating process was applied at $V_G$ = 4.0 V, $T$ = 250 K, for 30 minutes. The root-mean-square roughness over the whole 5 μm by 5 μm area is ~0.12 nm, ~0.08 nm, and ~0.14 nm for (**A**), (**B**) (excluding the dust at the right bottom corner), and (**C**), respectively.

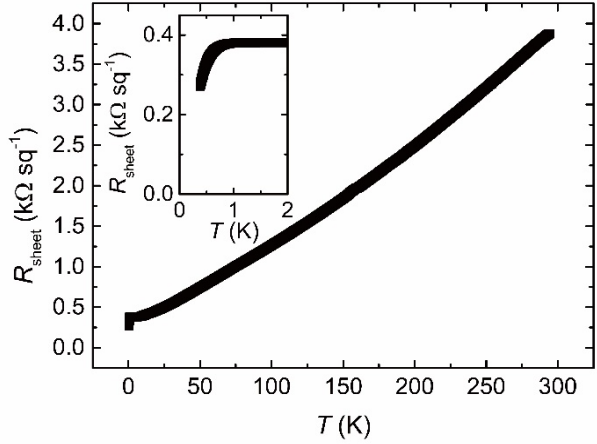

**Fig. S3. Temperature-dependent $R_{sheet}$ of an a-LAO(20-nm)/KTO(111) interface.** The interface has a metallic conduction over a wide temperature range, and exhibits a superconducting hint below $T = 1$ K. We note that in our IL-gated devices the amorphous LAO layer was used to ensure better electrical contacting to the central Hall-bar channel (Fig. 1**D**). The superconductivity at this a-LAO/KTO interface is irrelevant with the superconductivity we achieved in the channel by IL gating.

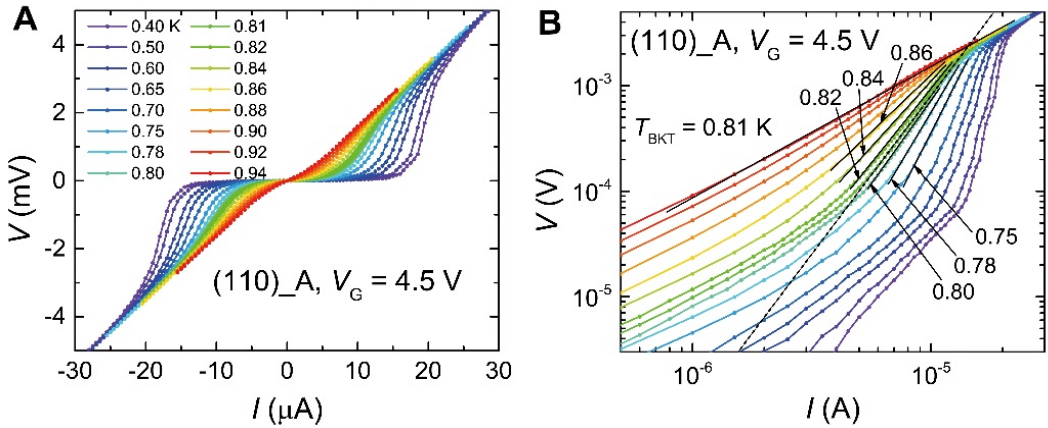

**Fig. S4. Voltage-current (*V-I*) measurements for IL-gated (110) surface.** (**A**) Temperature-dependent *V-I* characteristics of the (110)_A sample at $V_G = 4.5$ V. The channel width is 100 μm. (**B**) the same *V-I* curves on a logarithmic scale. The long dashed line corresponds to $V \propto I^3$ dependence and shows that $T_{BKT} = 0.81$ K.

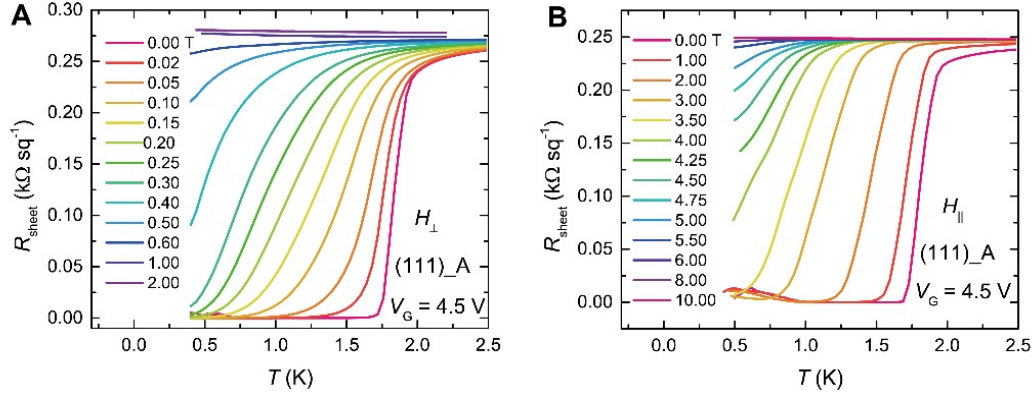

**Fig. S5. Sheet resistance $R_{sheet}$ of (111)_A gated at $V_G$ = 4.5 V under different magnetic fields.** (**A**) For fields perpendicular to the surface. (**B**) For fields parallel to the surface.

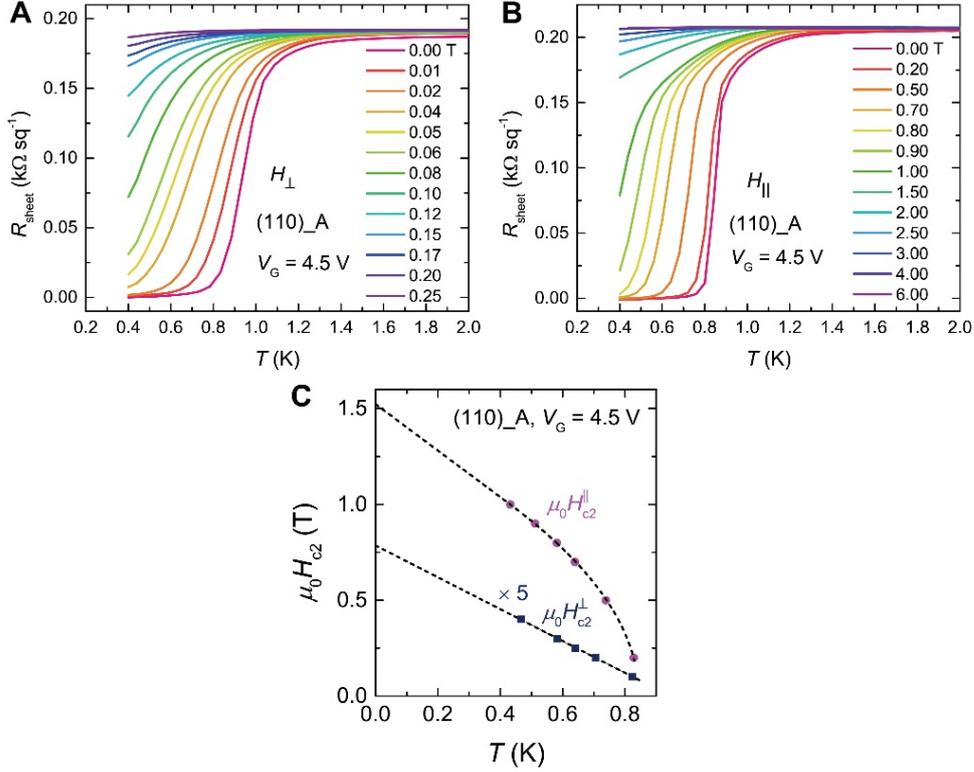

**Fig. S6. Sheet resistance $R_{sheet}$ of (110)_A gated at $V_G$ = 4.5 V under different magnetic fields.** (**A**) For fields perpendicular to the surface. (**B**) For fields parallel to the surface. (**C**) The upper critical field $\mu_0 H_{c2}$ versus temperature. The $\mu_0 H_{c2}^{\perp}$ (squares; field data multiplied by 5 for clarity) and $\mu_0 H_{c2}^{\parallel}$ (circles) label the upper critical field perpendicular and parallel to the sample surface respectively. Dashed lines are fits to linearized Ginzburg-Landau theory.

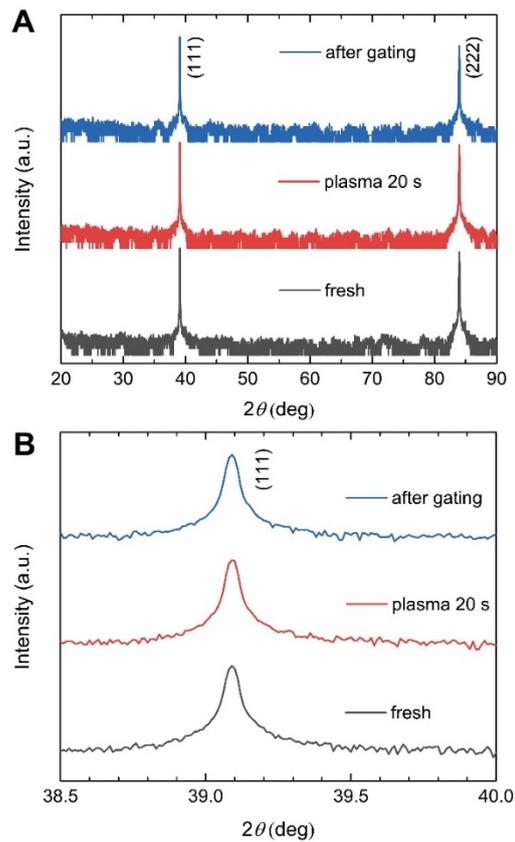

**Fig. S7. Out-of-plane *θ*-2*θ* X-ray diffraction on a KTO(111) substrate surface in three different situations.** The definition of the three situations is the same as that described in the caption of Fig. S2. (**A**) In a wide 2*θ* range. (**B**) In a close view around the KTO(111) peak. No detectable change was observed in the three different situations.